\documentclass[a4paper, twocolumn, 11pt, unpublished]{quantumarticle}
\pdfoutput=1
\usepackage[utf8]{inputenc}
\usepackage[english]{babel}
\usepackage[T1]{fontenc}
\usepackage{amsmath, amsfonts, amssymb, amsthm}
\usepackage{physics, mathtools}
\usepackage{breakcites, microtype}
\usepackage[breaklinks=true]{hyperref}
\usepackage{colortbl}

\DeclareFontFamily{U}{mathx}{\hyphenchar\font45}
\DeclareFontShape{U}{mathx}{m}{n}{
      <5> <6> <7> <8> <9> <10>
      <10.95> <12> <14.4> <17.28> <20.74> <24.88>
      mathx10
      }{}
\DeclareSymbolFont{mathx}{U}{mathx}{m}{n}
\DeclareFontSubstitution{U}{mathx}{m}{n}
\DeclareMathSymbol{\bigtimes}{1}{mathx}{"91}

\usepackage{tikz}
\usepackage{lipsum}

\newtheorem{theorem}{Theorem}

\begin{document}

\title{Classical-quantum network coding: a story about tensor}

\author{Clément Meignant}
\affiliation{Sorbonne Universit\'e, CNRS, LIP6, F-75005 Paris, France}
\orcid{}
\author{Frédéric Grosshans}
\orcid{0000-0001-8170-9668}
\affiliation{Sorbonne Universit\'e, CNRS, LIP6, F-75005 Paris, France}
\author{Damian Markham}
\affiliation{JFLI, CNRS, National Institute of Informatics, University of Tokyo, Tokyo, Japan}
\affiliation{Sorbonne Universit\'e, CNRS, LIP6, F-75005 Paris, France}
\orcid{0000-0003-3111-7976}
\maketitle

\begin{abstract}
    We study here the conditions to perform the distribution of a pure state on a quantum network using quantum operations which can succeed with a non-zero probability, the Stochastic Local Operation and Classical Communication (SLOCC) operations.
    In their pioneering 2010 work, Kobayashi et al.\@ showed how to convert any classical network coding protocol into a quantum network coding protocol. However, they left open
        whether the existence of a quantum network coding protocol implied the existence of a classical one.
    Motivated by this question, we 
    characterize the set of distribution tasks achievable with non zero probability
    for both classical and quantum networks. 
    We develop a formalism which encompasses
    both types of distribution protocols
    by reducing
    the solving of a distribution task to the factorization of a tensor
     in $\mathbb C$ or $\mathbb R^+$. Using this formalism, we examine the equivalences and differences between both types of distribution protocols exhibiting
    several elementary and fundamental relations between them as well as concrete examples of both convergence and
    divergence. We answer
    by the negative to the 
    issue previously left open: some tasks are achievable in the quantum setting, but not in the classical one. 
    We believe this formalism to be a useful tool for studying the extent of quantum network ability 
    to perform multipartite distribution tasks. 
\end{abstract}
\section{Introduction}

\begin{figure}[tb]
    \centering
\includegraphics[width=0.7\columnwidth]{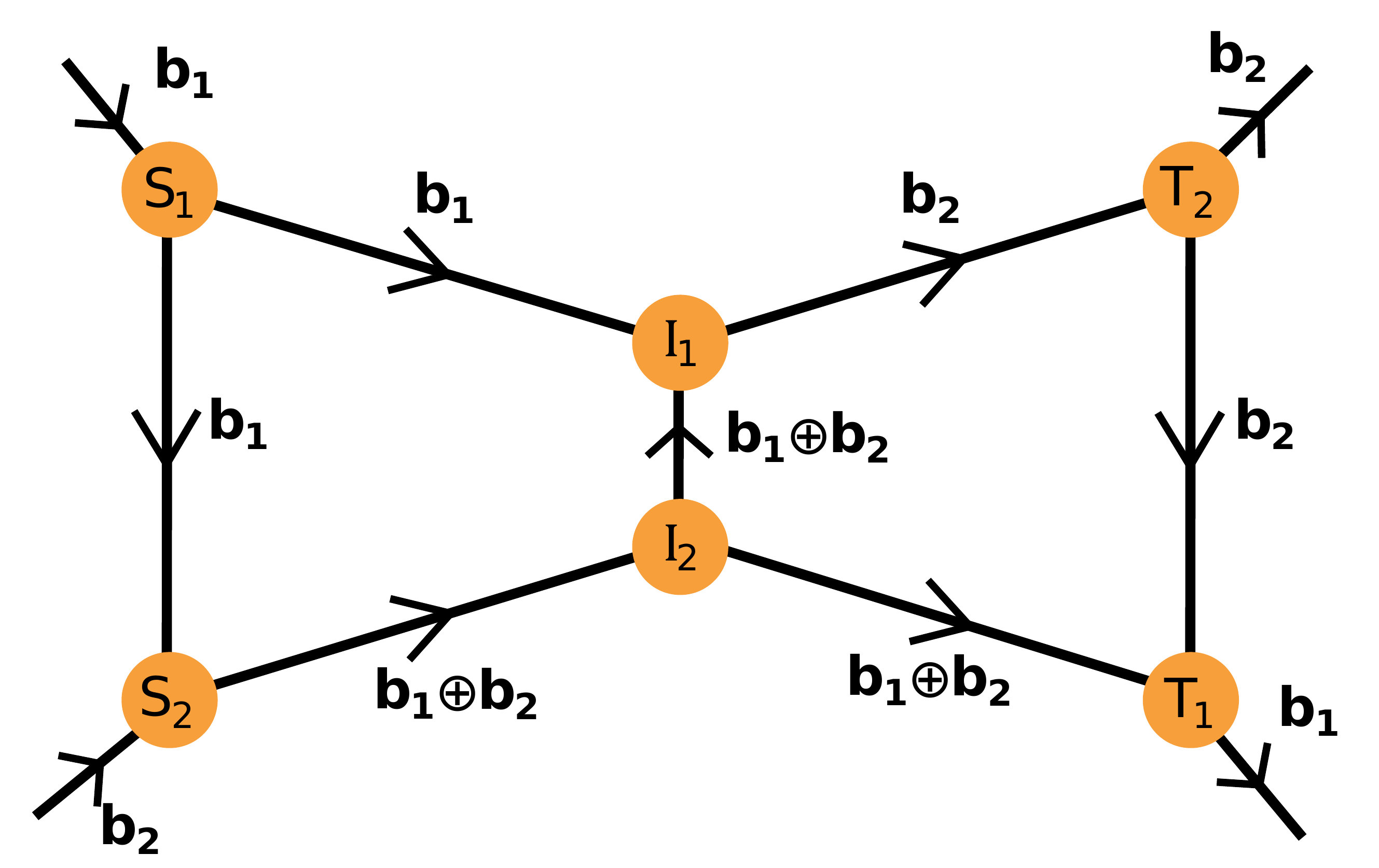}
    \caption{ Classical network coding over the butterfly network. Here, routing a bit from $S_2$ to $T_2$ and from $S_1$ to $T_1$ with a single channel use is impossible. But by sending the addition modulo $2$ through the central channel allow the distribution to be performed.\label{fig:butterfly}}
\end{figure}

\begin{figure}[tb]
    \centering
\includegraphics[width=1.\columnwidth]{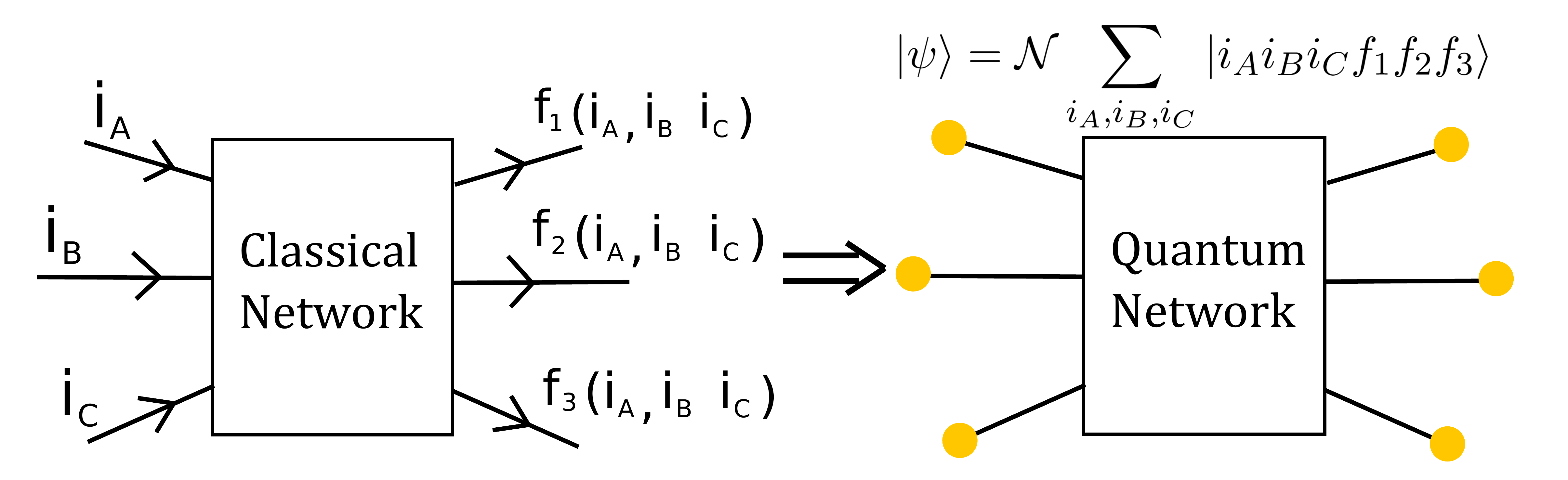}
    \caption{Classical network coding allows to put into communication several clients of the network. In the most general sense, some clients input some bits ($i_A$, $i_B$ and $i_C$ on the left side) and other clients receive arithmetical functions of those bits ($f_1$, $f_2$ and $f_3$ on the right side). For a given topology of network, if there is a classical network coding scheme such that by inputing $\vec{i} := (i_1,\dots, i_n)$ the network outputs $\vec{f}(\vec{i}) := (f_1(\vec{i}),\dots,f_m(\vec{i}))$, you can distribute the pure state $\sum_{\vec{i}}\ket{\vec{i}\vec{f}}$ on the quantum network presenting the same topology. } \label{fig:classical_quantum}
\end{figure}

Quantum networks are currently a major investigation field in quantum technologies
\cite{wehner2018quantum}, with ongoing research on their theoretical characterization, the specifications of protocols to use them and the actual building of 
hardware to implement them.
Among possible applications of quantum networks are re implementation of
quantum protocols such as sensing \cite{degen2017quantum}, distributed computing \cite{briegel2009measurement} or quantum voting \cite{bao2017quantum} between
geographically distant parties; quantum networks are also useful at shorter range, inside a quantum computer, e.g.\@ 
to distribute resource state among several quantum processors or in
measurement based quantum computing \cite{briegel2009measurement}. 
Yet, all those applications need to use
different types of resource states which need to be distributed on the network. In the advent of quantum networking, it is necessary to examine all possibilities available to solve the issue of actually handling the distribution of resource states for both
communication and computation protocols. Moreover, the use of a global network by multiple users implies that simultaneous tentatives of communication will arise. In a quantum network, it
involves the simultaneous distributions of several states uncorrelated to each others and so, the distribution of product states. However, simply routing informations is sometimes impossible because of a lack of channel capacity. This issue can be solved in classical networks
by increasing greatly the channels capacity. This solution is not satisfying in the quantum case since increasing the capacity of a quantum channel is much more costly. The other solution, which have our preference, is to optimize the use of each channel and
thus study simultaneous distributions beyond simple routing. It is an problem we particularly wish to study as being able to perform simultaneous distributions is a feature essential to a common network such as the ``Quantum Internet'' \cite{kimble2008quantum}.\newline
The issue we solve here is the following: given a network's topology, which quantum states can be distributed with only Local quantum Operations and free Classical Communications, the LOCC class of operations? 
Kobayashi et al.\@ \cite{kobayashi2010perfect} introduced a method of distribution they named quantum network coding. It is based on classical network coding, a class of classical distribution protocols which includes routing but allows the nodes of a network to perform arbitrary arithmetic operations on their
inputs before outputing them.
This class of protocol allows to perform distribution tasks impossible for routing protocols, as shown by the canonical example of the
butterfly network (Figure \ref{fig:butterfly}). Kobayashi et al.\@ proved that, for each classical network coding distribution protocol, 
one could construct a 
protocol using Local Operations and Classical Communications (LOCC) to distribute the corresponding quantum state over the quantum counterpart of the network (Figure \ref{fig:classical_quantum}).
They left open the question of the existence of states which could be distributed by a LOCC quantum protocol with no classical equivalent.

Here, we study the full extent of quantum networks' distribution capability and compare it \textit{mutatis mutandis} to the classical one. First, we chose to consider stochastic LOCC (SLOCC), meaning we allow protocols to fail and consider them a success if they can succeed with a non zero probability. We
develop a new formalism inspired from the tensor network representation \cite{evenbly2011tensor} which reduces quantum state distribution to  tensor factorization and show a direct practical application by solving, using simple linear algebra, an
elementary yet difficult issue which is to decide whether a cross communication of two qubits can be achieved over a square shaped quantum network. Then, we show we can also reduce classical distribution over classical network as tensor factorization. The main difference being that classical network coding is only described with non-negative tensor. 
For the comparison of both settings to make sense, we restrict
our study to quantum subset states \cite{grilo2016qma} -- i.e.\@ states which can be described, up to local unitary operations, by  real
positive coefficients. This family of states encompass highly relevant states for quantum application such as Greenberger--Horn--Zeilinger (GHZ)-states, W-States or any bicolorable graph states \cite{briegel2001persistent}. From our formalism, we find a large set of quantum distribution protocols directly resultant from classical ones, it also becomes naturally clear we can find
quantum network coding protocols with no classical equivalent. Thus, we answer the previously open question negatively by giving a counter-example. This article is organized as follow: in Section \ref{sec:intro} we introduce the general formalism which
encapsulates a coding setup and a distribution task. In section \ref{sec:qnc}, we present the extended version of quantum network coding, which represent the capabilities of quantum networks to solve distribution tasks and we show that the solving of a specific task is equivalent to the finding of a tensor's
factorization. We show results which exploit this formalism to prove the impossibility to perform a simultaneous quantum communication over a square quantum network. In section \ref{sec:cnc}, we define classical network coding and the class of distribution tasks it can solves. Finally, we show that both quantum network coding and classical network coding can be
unified under the same tensor formalism in section \ref{sec:result}, the difference being that quantum network coding allows the factorization to be done with tensors in $\mathbb{C}$ while 
for classical network coding it is restricted to non-negative tensors, tensor with coefficients in
$\mathbb{R}^+$. Then, we show a direct application of our formalism to find the minimal resources needed to perform simultaneous communication over a square shaped network. We
conclude this paper by giving an example of a distribution task achievable through quantum network coding while impossible with
classical network coding, solving the converse of the problem arisen by Kobayashi et al.\@ and expanding the extent of the full capabilities of quantum networks.

\section{Definitions}
\label{sec:intro}
Whether it is classical or quantum, we model a network as a set of nodes $v$ linked by perfect channels $e$ allowing the exchange of (qu)dits of dimension $d_e$. Formally, the network is
defined as the weighted graph $G=(V,E,D_E)$: $V$ is the set of all nodes, $E\subset V^2$ is the set of all channels and $D_E$ is the set of the channels' dimension. For $v\in V$, we call $N_v$ the neighbourhood of $v$ which is the set of all nodes sharing a channel with $v$.\newline
As said above, we wish to study which distribution tasks are possible over a given network. Distribution tasks are of disparate natures whether one consider the quantum or the
classical setup: in the first case, the task is the distribution of a specific quantum state; in the latter case, it is a distribution of classical correlations over the network. Nevertheless, we define a general formalism encompassing both cases.
To define a distribution task, we first need to know its set of clients, which are
the nodes sharing the final correlation after distribution.
Without loss of generality\footnote{A client leaf can be formally added for non-leaf clients, and non-client leaves can safely be ignored and removed from the network.}, the leaves of the network --- the nodes of degree $1$ ---
will be the clients and denote the set of all coding clients as $C\subset V$.
The tasks themselves are defined below, first for the quantum setup, and then for the classical one.

\section{Quantum Network Coding}
Let $G=(V,E,D_E)$ be a network in the quantum setting, and $C\subset V$ the set of clients. 
In our setup where classical communication is free, a perfect channel is equivalent to a maximally entangled states of the same dimension. As a consequence, each node $v\in V$ will own a part of a
maximally entangled state (also known as EPR pairs\footnote{Named after Einstein Podolsky and Rosen’s  seminal article \cite{EPR35}.} or Bell pairs) for each incident edge. We label each qudit and denote the qudit belonging to edge $e$ of node $v$ as subsystem $v_e$. We describe the initial (unnormalized) state of the quantum network as
\begin{equation}
    \ket{G}^{V} \propto
        \sum_{\overrightarrow{i_E}}\bigotimes_{e\in E}\ket{i_ei_e}^{v_e w_{e}}
\end{equation}
where $\overrightarrow{i_E} := (i_e)_{e\in E}$.
A distribution task corresponds to the distribution of a pure state $\ket{\psi}^C$ shared among the clients in $C$. Such a task is possible over 
the network with nonzero probability iff
one can find a stochastic LOCC (SLOCC) protocol transforming $\ket{G}^V$ into $\ket{\psi}^C$.
The \emph{stochastic} part is crucial as we wish to characterize the full 
set of states a quantum network can distribute with nonzero success probability.
Formally, distribution tasks are described as target states in the computational basis by
\begin{equation}
    \ket{\psi}^{C} = \sum_{\overrightarrow{i_C}}\mathcal{T}^{\overrightarrow{i_C}}\bigotimes_{c\in C}\ket{i_c}^c
    \label{eq:distribtaskT}
\end{equation}
where $\overrightarrow{i_C} := (i_c)_{c\in C}$. 
We notice we can fully characterize the distribution task by a tensor $\mathcal{T}^{\overrightarrow{i_C}}$ with coefficients in $\mathbb{C}$.
These coefficients are the ones of the target state $\ket{\psi}^C$,
expressed in the computational basis.
To extract the wanted conditions for conversion,
we first recall a necessary and sufficient condition condition for a SLOCC to convert 
the state $\ket{G}^V$ into $\ket{\psi}^C$ \cite{walter2016multipartite}.
Such a conversion is possible
if and only if there exists a LOCC protocol such that at least one branch achieves the
transformation. Formally, if a SLOCC can convert $\ket{G}^V$ to $\ket{\psi}^C$, we can find a set of Kraus operators $\{K_v\}_{v\in V}$ and $\lambda \in \mathbb{R}^{+*}$ such that
    $\bigotimes_{v\in V}K_v \ket{G}^V = \lambda\ket{\psi}^C$.
Equivalently, a SLOCC can convert $\ket{G}$ to $\ket{\psi}$ if and only if there exists a set of matrices $\{M_v\}_{v\in V}$ such that:
\begin{equation}
    \smashoperator{\bigotimes_{v\in V}} M_v \ket{G}^V = \ket{\psi}^C.
\end{equation}
Trivially, the set of matrices $\{M_v\}_{v\in V}$ can be converted into Kraus operators by a simple normalization. Writing the action of the matrices $M_v$ on the tensor network representation, we deduce 
the following necessary and sufficient condition for a distribution task 
$\mathcal{T}^{\overrightarrow{i_C}}$ --- as defined in eq.\ \eqref{eq:distribtaskT} ---
to be achievable over a quantum network.
\begin{theorem}
    \label{thm:qnc}
 Let $G=(V,E,D_E)$ be a quantum network, 
 a distribution task $\mathcal{T}^{\overrightarrow{i_C}}$ is achievable by quantum network coding over
 the network if and only if $\mathcal{T}^{\overrightarrow{i_C}}$ can be factorized as
 \begin{equation}
     \mathcal{T}^{\overrightarrow{i_C}} = \prod_{v \in V\setminus C} \mathcal{V}^{\overrightarrow{i_v}}
     \label{eq:state_description}
 \end{equation}
 where each $v\in V$ is associated to tensor $\mathcal{V}^{\overrightarrow{i_v}}\in\mathbb{C}^{\bigtimes_{w\in N_v} d_{\{v,w\}}}$ and tensors are contracted along indices corresponding to shared edges.
\end{theorem}
\begin{proof}
  Let us suppose $\ket{\psi}^{C}$ is achievable by a SLOCC protocol.
  Since we are only interested by the existence of such a protocol and not its actual probability of success
  as long as it is non null, we may assume, without loss of generality, this SLOCC
  is the tensor product of successful single-dimensional projective measurements on all non-client nodes $V\setminus C$.
  More formally, for each node $v\in V\setminus C$, there exists an unnormalized bra
  \begin{equation}
      \bra{\mathcal{V}_v}=\sum_{\overrightarrow{i_v}}\mathcal{V}_{\overrightarrow{i_v}}\bra{\overrightarrow{i_v}},
  \end{equation}
  where $\mathcal{V}_{\overrightarrow{i_v}}\in \mathbb C$ for all $\overrightarrow{i_v}:=(i_{v_e})_{\substack{e\in E}\\ v\in e}$
  such that
  \begin{equation}
  \ket{\psi}^C =
     \left( \smashoperator[r]{\bigotimes_{v\in V\setminus C}} \bra{\mathcal{V}_v}\right) \ket{G}^V. 
  \end{equation}
  We then have
  \begin{align}
     \ket{\psi}^C &= 
      \left(\smashoperator[r]{\bigotimes_{v\in V\setminus C}}\,\,\,    \sum_{\overrightarrow{i_v}}\mathcal{V}_{\overrightarrow{i_v}} \bra{\overrightarrow{i_v}}\right)\left( \sum_{\overrightarrow{i_E}}\bigotimes_{e\in E}\ket{i_ei_e}^{v_e w_{e}}\right)\\
      &= \sum_{\overrightarrow{i_E}}\left(\smashoperator[r]{\bigotimes_{v\in V\setminus C}}\,\,\,    \sum_{\overrightarrow{i_v}}\mathcal{V}_{\overrightarrow{i_v}} \bra{\overrightarrow{i_v}}\right)\left( \bigotimes_{e\in E}\ket{i_ei_e}^{v_e w_{e}}\right)\\
      &= \sum_{\overrightarrow{i_C}}\left(\smashoperator[r]{\bigotimes_{v\in V\setminus C}}\,\,\,    \sum_{\overrightarrow{i_v}}\mathcal{V}_{\overrightarrow{i_v}}\prod_{e\in E} \delta^{i_{v_e}}_{i_{w_e}}\right)\bigotimes_{c\in C} \ket{i_c} \\
      &= \sum_{\overrightarrow{i_C}}\prod_{v \in V\setminus C} \mathcal{V}^{\overrightarrow{i_v}}\bigotimes_{c\in C}\ket{i_c},
  \end{align}
 where in the last line, we sum implicitly over repeated indices.
 By identification, we observe the requested equality:
\begin{equation}
    \mathcal{T}^{\vec{i_v}} = \prod_{v} \mathcal{V}^{\vec{i_v}}
\end{equation}
Conversely, if the equality is verified we can extract projective operators distributing $\ket{\psi}$. 
\end{proof}

This theorem is already a powerful tool in order to study distribution over a quantum network. We already mentioned our focus on multi-user network. From those networks arise
the need to be able to handle simultaneous distributions of states while minimizing the consumption of entanglement. As a consequence, we chose to study an elementary and critical example, the distribution of two EPR pairs over a network. As
mentioned in the introduction, it was already proven that this distribution task was achievable over a butterfly network \cite{kobayashi2010perfect}. However, a question lingered, can we perform a cross distribution over a smaller network, namely a butterfly network which is missing an intermediary channel?\newline
The only existing proof \cite{akibue2016network} which tackled this question is quite involved and difficult to generalize. We show here that,
using theorem \ref{thm:qnc}, only basic linear algebra is required to show that, as in the classical case, cross-distribution remains impossible to perform over a square
network.
\begin{proof}
  Let $G$ be the network defined and let $\mathcal{T}$ be the distribution task of a cross distribution of EPR over this network as in Figure \ref{fig:x_box2}
\begin{figure}
    \centering
\includegraphics[width=0.70\columnwidth]{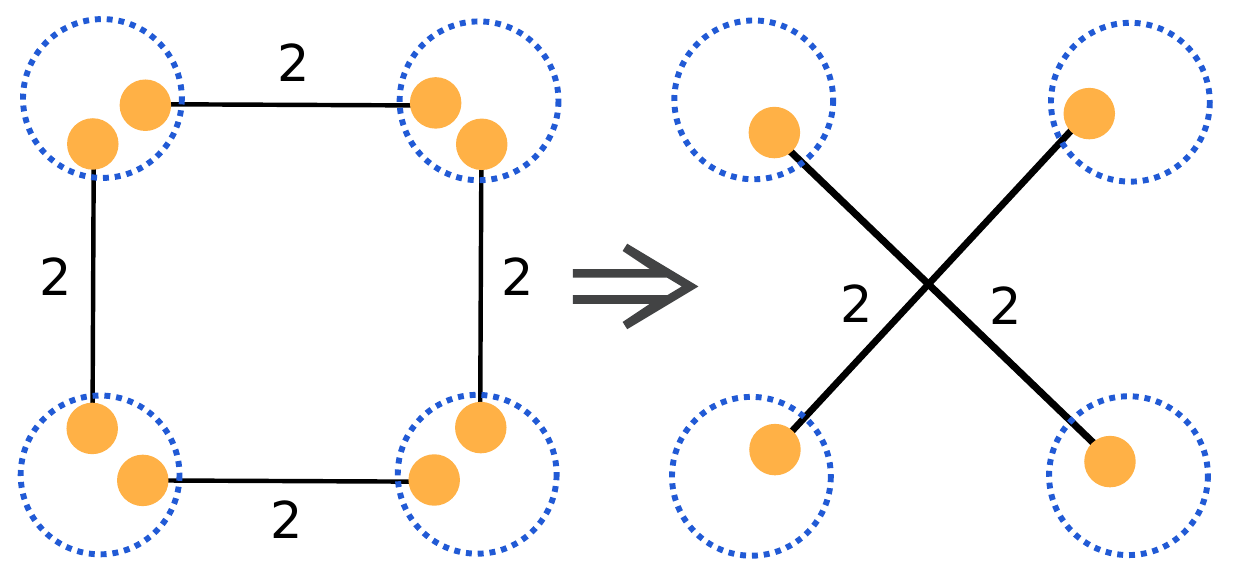}
    \caption { The distribution task we wish to achieve: distributing cross-EPR pair across a square network.\label{fig:x_box2}}
\end{figure}
\label{sec:XBoxProf}
We want to find a factorization of $\mathcal{T}^{i_A,i_B,i_C,i_D} = \delta_{i_A,i_C}\delta_{i_B,i_D}$ over the square network, which means finding $\mathcal{A}$, $\mathcal{B}$, $\mathcal{C}$, $\mathcal{D} \in \mathbb{C}^{2\times 2 \times 2}$ such that
\begin{equation}
    \mathcal{T}^{i_Ai_Bi_Ci_D} = \mathcal{A}^{i_A\alpha\beta}\mathcal{B}^{i_B\beta\gamma}\mathcal{C}^{i_C\gamma\delta}\mathcal{D}^{i_D\delta\alpha},
\end{equation}
using Einstein notation as we sum implicitely on repeated indices.
Namely, we want to find $\mathcal{A}^{i_A}$, $\mathcal{B}^{i_B}$, $\mathcal{C}^{i_C}$ and $\mathcal{D}^{i_D}$ in $\mathbb{C}^{2\times 2}$, which we represent as solution set
\begin{equation} \label{eq:condset}
\begin{vmatrix}
    \mathcal{A}^{0} & \mathcal{B}^{0} & \mathcal{C}^{0} & \mathcal{D}^{0} \\
    \mathcal{A}^{1} & \mathcal{B}^{1} & \mathcal{C}^{1} & \mathcal{D}^{1}
\end{vmatrix}
\end{equation}Such that 
\begin{equation}
    \Tr(\mathcal{A}^{i_A}\mathcal{B}^{i_B}\mathcal{C}^{i_C}\mathcal{D}^{i_D}) = \delta_{i_A,i_C}\delta_{i_B,i_D}
\label{eq:cond}
\end{equation}
There is several necessary conditions for the existence of solutions to equation (\ref{eq:cond}).
First one, the trace is a hermitian form. As a consequence, Eq.\@ (\ref{eq:cond}) implies that all families of consecutive matrices -- $\{\mathcal{A}^{i_A}\mathcal{B}^{i_B}\}$, $\{\mathcal{B}^{i_B}\mathcal{C}^{i_C}\}$, $\{\mathcal{C}^{i_C}\mathcal{D}^{i_D}\}$ and $\{\mathcal{D}^{i_D}\mathcal{A}^{i_A}\}$-- are bases of $\mathbb{C}^{2\times2}$.
The second one is that all matrices have to be of rank 2. 
Indeed, if at least one of the describing matrices is of rank 1, no solution can be found.
We focus now on the equation set in order to reduce the number of free parameters. 
First, we use a simple symmetry, generic to the trace:
for any invertible square  matrix  $M\in\mathbb{C}^{2\times2}$, we can replace the solution set (\ref{eq:condset}) by
\begin{equation}
\begin{vmatrix}
    \mathcal{A}^{0}M^{-1} & M\mathcal{B}^{0} & N\mathcal{C}^{0} & \mathcal{D}^{0}\\
    \mathcal{A}^{1}M^{-1} & M\mathcal{B}^{1} & N\mathcal{C}^{1} & \mathcal{D}^{1}
\end{vmatrix}
\end{equation}
since the trace is invariant under this transformation.
Moreover, it is easy to check that, for $\alpha \in \mathbb{C}^*$ and $\lambda \in\mathbb{C}$,
\begin{equation}
\begin{vmatrix}
    \mathcal{A}^{0} & \mathcal{B}^{0} & \mathcal{C}^{0} + \lambda\mathcal{C}^1 & \mathcal{D}^{0}\\
    \alpha(\mathcal{A}^{1} - \lambda\mathcal{A}^0) & \mathcal{B}^{1} & \alpha^{-1}\mathcal{C}^{1} & \mathcal{D}^{1}
\end{vmatrix}
\end{equation}
is also solution thanks to symmetries of the solution set. Finally, the cyclic property of the trace is the last symmetry we exploit. The use of all of these symmetries imply the existence of a solution of the form
\begin{equation}
\begin{vmatrix}
    \mathbb{I} & \mathbb{I} & \mathcal{C}^{0} & \mathbb{I} \\
    \mathcal{A}^{1} & Z & \mathcal{C}^{1} & \mathcal{D}^{1}
\end{vmatrix},
\end{equation}
with $\mathcal{A}^1_{0,0} = 1$ ---we used previous symetries with $\alpha$ being the inverse of the original coefficient---  which reduces to 15 the number of parameters to solve a set of 16 free equations. Thus, proving the non-existence of any solution to perform a cross distribution on a square quantum network.
\end{proof}\label{sec:qnc}

We have therefore reduced the problem of the distribution of a pure quantum-state by SLOCCs to a tensor factorization, and have 
used this result to prove the impossibility of a practical information task. We will now focus on the study of classical communications over classical networks to extract similar kind of conditions.

\section{Stochastic Classical Network Coding} 
\label{sec:cnc}
In order to
be able to compare fairly the quantum and classical settings, we now adapt the formalism developed above to classical network coding.
In a classical distribution task,
clients are partitioned between the inputs---the sources $S\in C$---and the outputs---the sinks $T\in C$---of the
network. 
In the literature \cite{koetter2003algebraic,
ho2008network}, network coding protocols are typically 
deterministic, i.e.\  each input $\vec{i} := (i_c)_{c\in S}$ at the
sources always yields a specific output $\vec{o}:=(i_c)_{c\in T}$ at
the sinks.
However, we consider here the more general case of stochastic network
coding; these protocols are probabilistic and, like the SLOCCs studied in
the previous question, can abort with non zero probability. 
Formally, a distribution task is
defined by a set of probabilities, for each input $\vec{i}$, the network outputs
$\vec{o}$ with probability $p(\vec{o}|\vec{i})$ and the protocol abort with probability $1 - \sum_{\vec{o}} p(\vec{o}|\vec{i})$. We gather all probability in a tensor
$\mathcal{T}^{\vec{i},\vec{o}} = p(\vec{o}|\vec{i})$ with coefficient in $\mathbb{R}^+$. 
As the sets of sources and sinks form together a partition of the set of clients, $\mathcal{T}^{\vec{i},\vec{o}}$ can be written as
the tensor
$\mathcal{T}^{\overrightarrow{i_C}}$, similarly to eq.\ \eqref{eq:distribtaskT} --which defines the quantum setting--.
One can therefore defines a classical and a quantum distribution task using
the same formalism, the only difference up to normalization between the tasks being that the classical tensor is restricted to real non-negative 
coefficients while the quantum one can also have negative and complex coefficients.
To study the relations between
distributions in the two setups, we need to restrict the distribution task we consider in the quantum one to ones which are described by non-negative tensors. Namely, the distribution of quantum subset states.\newline
Using the, now introduced, tensor formalism, we characterize necessary and sufficient conditions for a distribution task to be achievable over a given classical network.
 We can extend a bit this definition by noticing that multiplying $\mathcal{T}$ by a strictly positive scalar $\lambda \in \mathbb{R}^{+*}$ will only alter the general probability of
 success without changing the correlation between inputs and outputs. As a consequence, we deem that a distribution task $\mathcal{T}$ is achievable on a classical network if there
 exists a $\lambda$ strictly positive such that  $\lambda\mathcal{T}$ is achievable.\newline
How does one implement a Stochastic classical network coding to perform a distribution task? The answer is quite straightforward: in a network coding setup,
each node waits inputs $\vec{i_v}$, then performs an arithmetical operation such
that it outputs $\vec{o_v}$ with probability
$p_v(\vec{o_v}|\vec{i_v})$. As before, we regroup the probability table in tensor $\mathcal{V}^{\vec{i_v},\vec{o_v}} = p_v(\vec{o_v}|\vec{i_v})$. We can compute the
tensor given by two nodes sharing a link: let $v$ and $w$ be two nodes of the network such that some outputs of $v$ are inputs of $w$. We call $\vec{i_v}$ the inputs of $v$, $\vec{j}$
the inputs of $w$ which are output of $v$, $\vec{o_v}$ the other
outputs of $v$, $\vec{i_w}$ the inputs of $w$ which are not output of $v$ and, finally, $\vec{o_w}$ the outputs of $w$. The probability to input $(\vec{i_v},\vec{i_w})$ and output $(\vec{o_v},\vec{o_w})$ is given by:
\begin{equation}
    p(\vec{o_v},\vec{o_w}|\vec{i_v},\vec{i_w}) = \sum_{\vec{j}} p_v(\vec{o_v},\vec{j}|\vec{i_v}).p_w(\vec{o_w}|\vec{j},\vec{i_w})
\end{equation}
The tensor relation becomes relevant here, grouping the probability table $p$ in a tensor $\mathcal{P}^{\vec{o_v},\vec{o_w},\vec{i_v},\vec{i_w}} = p(\vec{o_v},\vec{o_w}|\vec{i_v},\vec{i_w})$, we can rewrite the previous equation as a contraction of tensors:
\begin{equation}
    \mathcal{P}^{\vec{o_v},\vec{o_w},\vec{i_v},\vec{i_w}} = \mathcal{V}^{\vec{o_v},\vec{j},\vec{i_v}}\mathcal{W}^{\vec{o_w},\vec{j},\vec{i_w}}
\end{equation}
where we used the Einstein notation to sum implicitly over repeated indices.\newline
So, each stochastic classical network coding  is an assignation to each node $v$ of a non-negative tensor $\mathcal{V}\in {\mathbb{R}^+}^{\bigtimes_{w\in N_v} d_{\{w,v\}}}$, where there is an index for each incident channel and of the dimension of said channel. The distribution task associated to the assignation is the
contraction of those tensors along the network's edges. Everything is set for the subsequent theorem:
\begin{theorem}
\label{thm:scnc}
    Let $G=(V,E,D_E)$ be a classical network, a distribution task $\mathcal{T}^{\vec{i_c}}$ is achievable by a stochastic classical network coding over the network if and only if $\mathcal{T}^{\vec{i_c}}$ can be factorized as
 \begin{equation}
     \mathcal{T}^{\vec{i_c}} = \prod_{v \in V} \mathcal{V}^{\vec{i_v}}  \label{eq:state_description_class}
 \end{equation}
 where each $v\in V$ is associated to tensor $\mathcal{V}^{\vec{i_v}}\in{\mathbb{R}^{+}}^{\bigtimes_{w\in N_v} d_{\{v,w\}}}$ and tensors are contracted along indices corresponding to shared edges.
\end{theorem}

\section{Results} \label{sec:result}
As the main difference here is that classical network coding is only described with non-negative tensor. In order to keep the comparison between the quantum and the classical task meaningful, we restrict our study to subset states \cite{grilo2016qma}, i.e.\@ states described by real positive coefficients in the computational basis, up to local unitary operations. This family of states encompass highly relevant states for quantum applications such as GHZ-states, W-States or any bicolorable graph states.
By unifying theorem \ref{thm:qnc} and \ref{thm:scnc}, we can extract the following theorems
\begin{theorem}
 Any distribution task $\mathcal{T}$ achievable on a network by a stochastic classical network coding protocol is achievable on the corresponding quantum 
 network by a quantum network coding protocol.  \label{thm:corres}
\end{theorem}
\begin{theorem}
The converse is not true. \label{thm:nocnv}
\end{theorem}
\begin{proof}
The implication from classical to quantum is the direct consequence of both Theorems \ref{thm:qnc} and \ref{thm:scnc}: if a tensor is factorizable along the network topology in
$\mathbb{R}^+$, the tensor is factorizable in $\mathbb{C}$. We previously proved the impossibility of a cross distribution over
a square network with link of dimension 2. Applying theorem \ref{thm:corres}, we show that cross distribution is achievable using ternary channels and we go further by showing a LOCC exists --see Appendix \ref{ap:3to2}--, which means the operation can succeed with probability one,
a feat that cannot be achieved with classical network coding. This show that using ternary channels over quantum networks leads to a superiority into solving bottleneck issues.\newline
We now prove Theorem \ref{thm:nocnv}, and show that the question left-open by Kobayashi et al.\@ \cite{kobayashi2010perfect} can be answered negatively. 
As shown above, distributing $\ket{\psi}$ means finding a
factorization of $\mathcal{T}$ in the field $(\mathbb{C}, +, \cdot)$, while achieving the probability table $p(\vec{o}|\vec{i})$ by a
stochastic classical network coding protocol is equivalent to the finding of a factorization of $\mathcal{T}$ in the semiring
$(\mathbb{R}^+, +, \cdot)$. 
A counterexample -- a tensor $\mathcal{T}$ and a network $G$
such that $\mathcal{T}$ can be factorized in $\mathbb{C}$ and not in $\mathbb{R}^+$
-- suffices to prove Theorem \ref{thm:nocnv}. We exhibit below such a counterexample, based on the noisy typewriter channel well known
in classical information theory \cite{shannon1948mathematical}. This is a channel defined on
a finite alphabet $(a_0,\dots,a_n)$ which transmits,
with probability $1/2$ each, either the input letter or the next letter in the alphabet. For example, taking alphabet ${0,1,2,3}$ the channel gives the following correlation table:
\begin{equation}
    \mathcal{T} = \frac{1}{2}\begin{pmatrix}
        1 & 1 & 0 & 0\\
        0 & 1 & 1 & 0 \\
        0 & 0 & 1 & 1 \\
        1 & 0 & 0 & 1 
    \end{pmatrix}.
\end{equation}
Viewed as a quantum network distribution task, it becomes
\begin{align}
    \ket{\psi}= \sum_{j=0}^{3}\ket{j}\otimes (\ket{j\oplus 1} + \ket{j})
\end{align}
up to normalization.\newline
The network we examine is composed of a single channel shared by two clients. We know the minimum dimension necessary to distribute $\ket{\psi}$, it is the rank of matrix $\mathcal{T}$,
which is 3. We can find two matrices $C\in \mathbb{C}^{4\times 3}$ and $F\in \mathbb{C}^{3\times 4}$ such that $\mathcal{T} = CF$, 
  as shown below
  \begin{align}
    C &= \begin{pmatrix}
            1 & 1 & 0 \\
            0 & 1 & 1 \\
            0 & 0 & 1 \\
            1 & 0 & 0
        \end{pmatrix}\\
    F &= \begin{pmatrix}
        1 & 0 & 0 & 1\\
        0 & 1 & 0 & -1 \\
        0 & 0 & 1 & 1
    \end{pmatrix},
\end{align}
Therefore, the 4-dimensional noisy typewriter channel can be compressed with nonzero success-probability in a ternary quantum channel.

Is the same feat achievable by a ternary classical channel? In other words, 

can we find two matrices of the same dimension in
$\mathbb{R}^+$ which factorize $\mathcal{T}$ ? We denote the $i^{\textrm{th}}$ line of $C$ as $\overrightarrow{c_i}$ and the
$j^{\textrm{th}}$ columns of $F$ ad $\overrightarrow{f_j}$. If $C$ and $F$ factorize $\mathcal{T}$, then $\mathcal{T}^{i,j} =
\overrightarrow{c_i}.\overrightarrow{f_j}$. It is easy to check  $\{ \vec{c_1}, \vec{c_2}, \vec{c_3}\}$ are free and thus form a basis of 
$\mathbb{C}^{1\times 3}$. As a consequence, $\exists(\lambda,\mu,\nu) \neq (0,0,0)$ such that $\vec{c_4} = \lambda \vec{c_1} + \mu \vec{c_2} + \nu \vec{c_3}$. The remaining part is straightforward:
\begin{equation*}
    \begin{array}{clc}
         &\vec{c_4}.\vec{f_1} = \lambda = \mathcal{T}^{4,1} = 1 & \\
         &\vec{c_4}.\vec{f_2} = \lambda + \mu = \mathcal{T}^{4,2} = 0 &\implies \mu = -1
    \end{array}
\end{equation*}
Thus, if the coefficient of $\vec{c_4}$ are in $\mathbb{R}^+$ then those of $\vec{c_2}$ are in $\mathbb{R}$.
This factorization is therefore not achievable in  $\mathbb{R}^+$, and this task is therefore only achievable in
a quantum setting, but not in a classical one.
\end{proof}

Further considerations can be done concerning the probability of
success achievable with quantum network coding if there exists a
deterministic classical network coding. By deterministic, we mean
that each input produce a specific output or is considered as an
error and leads to a failure. Such a protocol means we look for a
factorization of the distribution task $\mathcal{T}$ with tensors
such that $p(\vec{i}|\vec{o}) = 0$ or $1$. If we manage to find
such a classical protocol which succeed with probability $p$, we
can find a LOCC protocol distributing the corresponding quantum
state with probability $p$. The proof is only a short extension
of the Kobayashi et al.\@ proof. In this demonstration, the classical operations done at each node are simulated quantumly with a well chosen unitary operation. Here, we add a projection of the input qubit onto the space of the accepted inputs which lead to an identical probability to succeed.

\section{Conclusion}
We developed here a new formalism to study the distribution of
quantum states over quantum networks. This formalism is capable of
finding previously known results in a more simple way as the
impossibility to perform a cross distribution over a square shaped
quantum network and is useful to find new result, as the
possibility for quantum networks to perform the previously
considered cross distribution using ternary channel. A feat that
classical network coding cannot perform with probability one, this
knowledge prove the superiority of using ternary channel on
quantum networks in order to solve bottleneck issues on the
network. Moreover, in this formalism, showing the fundamental difference
between quantum and classical network coding became simple
as they were reduced as factorization of tensor in different sets.
Exposing a quantum distribution of a quantum subset states with
no classical equivalent was reduced to the problem of finding a
matrix with a non-negative rank superior to its rank, answering the question left open by Kobayashi et al. Moreover, by direct association, we can relate some classical
protocol with a non-zero probability to succeed with their quantum
equivalent, having the same probability of success. We have good hope that those results can benefit to the research about resource management over quantum networks and help to design optimized quantum networks.\newline
\section{Acknowledgments}
We thank Mohab Safey El Din and Jean-Charles Faugère for stimulating discussions. We thank Andrea Olivo for his numerical insights which gave us the hope needed to pursue our analytical investigations. 
We  acknowledge support  of  the  ANR  through  the  ANR-17-CE24-0035  VanQute project and from the European Union’s Horizon 2020 re-search and innovation program under grant agreement No 820445 QIA project.
\onecolumn\newpage
\appendix
\section{A direct application, achieving cross EPR pairs on a ternary network}
\label{ap:3to2}
\begin{figure}
    \centering
\includegraphics[width=0.40\columnwidth]{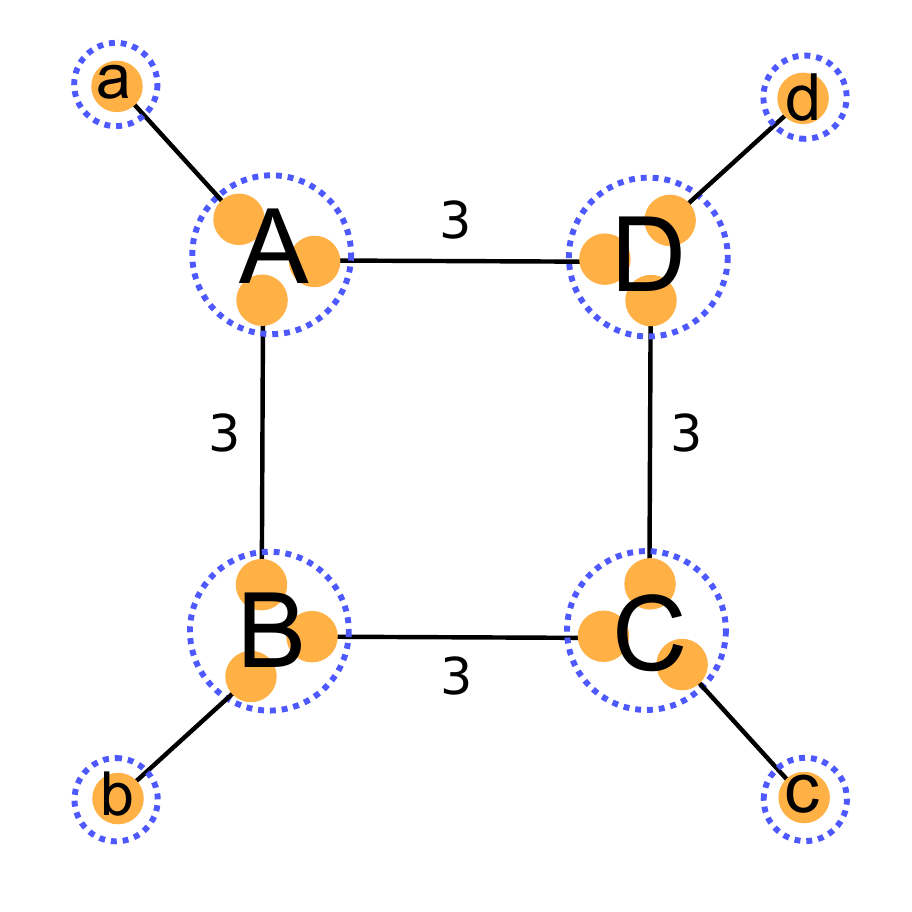}
    \caption { Graph defining the new network\label{fig:x_box3_square}}
\end{figure}

\begin{figure}
    \centering
\includegraphics[width=0.7\columnwidth]{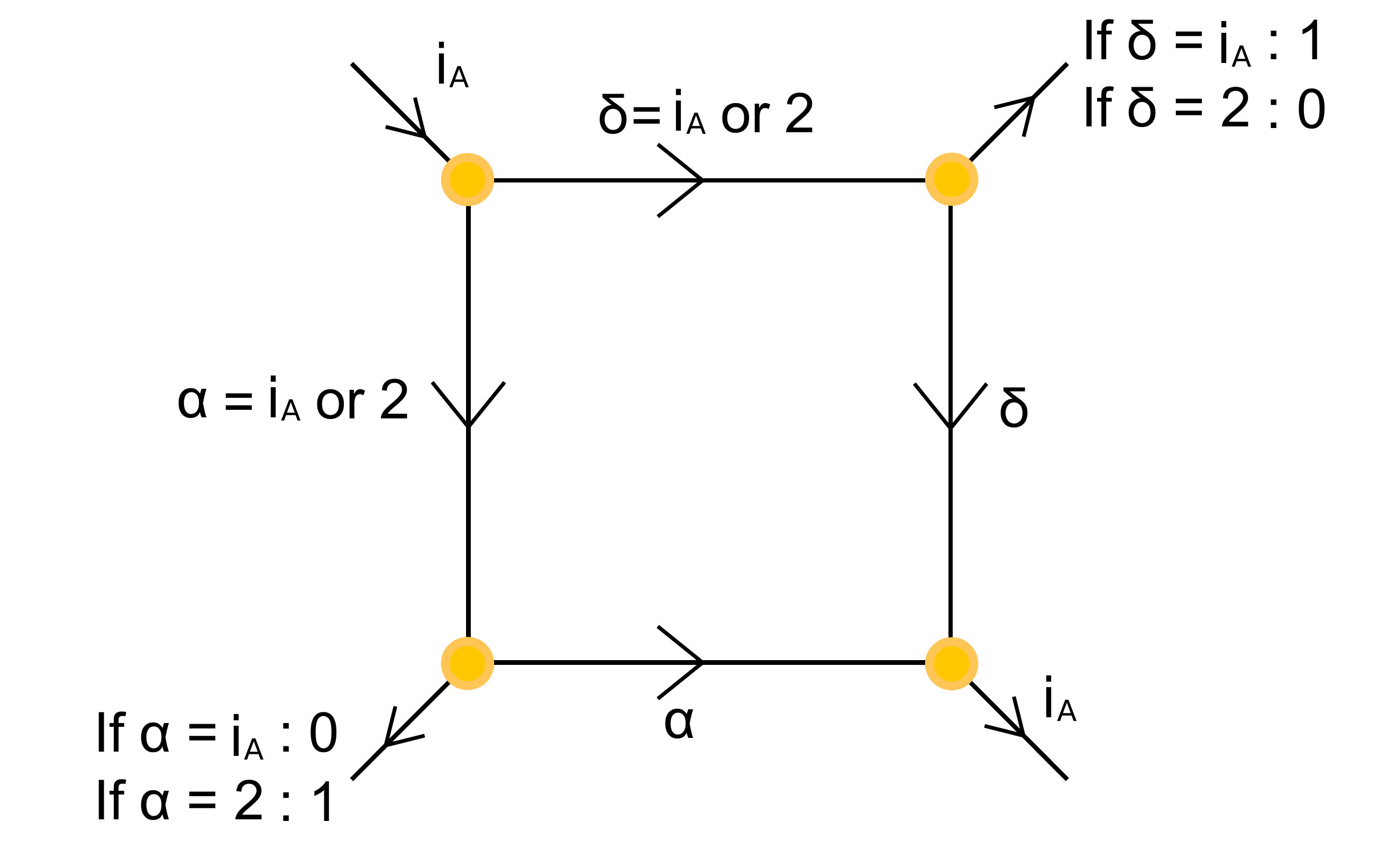}
    \caption {Classical scheme from which we can extract a LOCC to perfom a cross quantum communication over a square network. There is one source and 3 sinks here, the source will randomly decide which path will take $i_A$ and send on the network both $i_A$ and which path was taken, thus establishing the wanted correlation.\label{fig:x_box3}}
\end{figure}

\begin{figure}
    \centering
\includegraphics[width=1.\columnwidth]{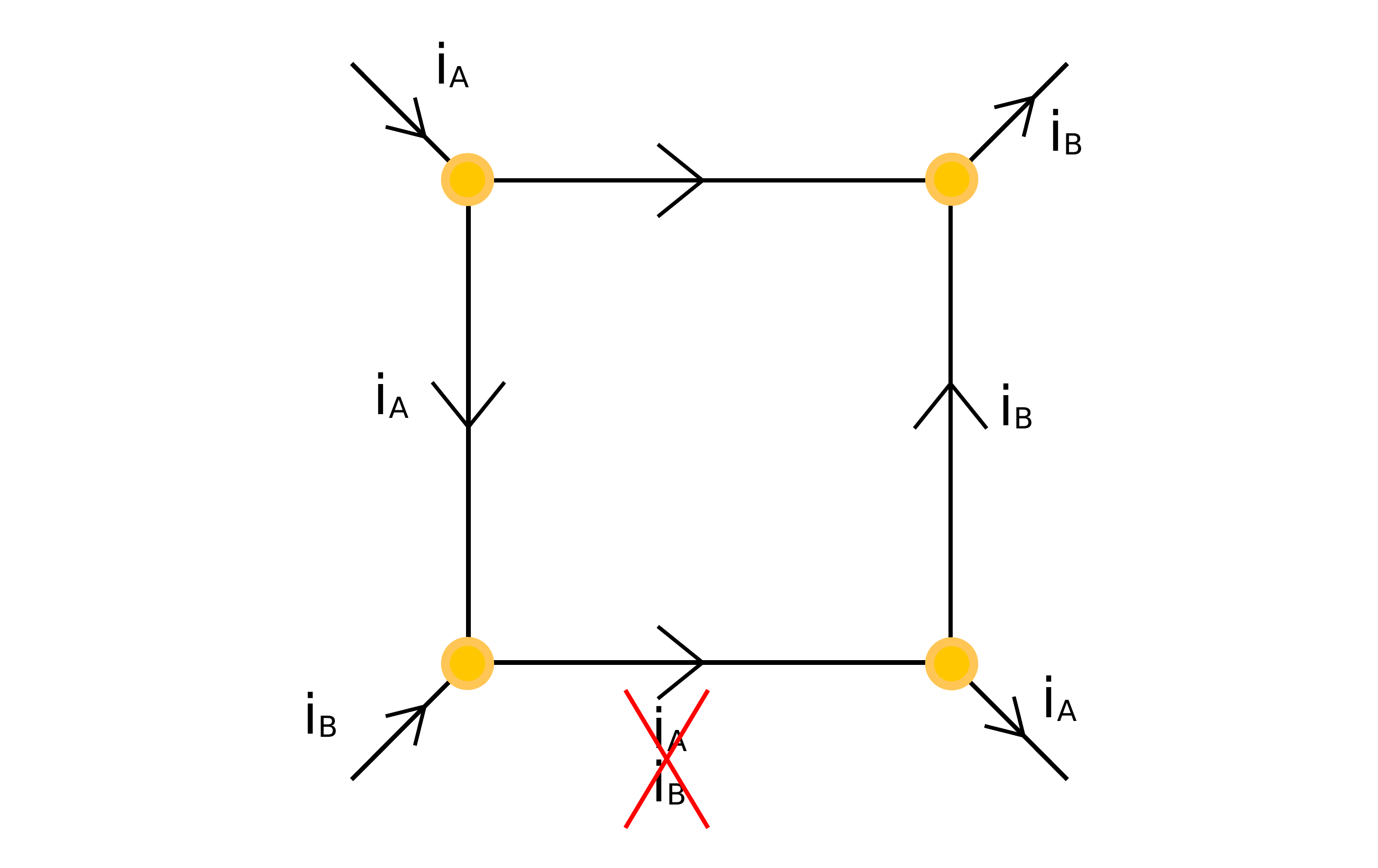}
    \caption {Cross communication is impossible to do classically with probability 1. Indeed, two bits of information have to go from the left-side to right side and two bits have to be exchanged between the top and the bottom. However, such choice of channels' direction involves that two bits have to pass into a ternary channel, which is impossible.\label{fig:class_imposs}}
\end{figure}

Let $G$ be a network defined by the graph depicted on Figure \ref{fig:x_box3_square}. We choose $a$, $b$, $c$ and $d$ as the clients. The correlation we wish to implement
is the following: we wish that $a$ and $c$ share a bit and that $b$ and $d$ share another uncorrelated bit. If we choose $a$ and $b$ as sources and $c$ and $d$ as sinks --a role's distribution expected if we look for cross distribution over the network--, no classical protocol can perform this distribution with probability 1, as can be seen in Figure \ref{fig:class_imposs}. However, choosing $a$ as the only source and the other clients as node, there is a protocol to
perform this distribution with probability 1, as can be seen in Figure \ref{fig:x_box3}\newline

We can extract from the classical protocol a quantum protocol allowing for the distribution of states $\ket{\psi} = \ket{\phi^+}^{AC}\otimes \ket{\phi^+}^{BD}$, where $\ket{\phi^+}$ is the maximally entangled pair of dimension 2, on the quantum network. The protocol is the following, A create locally the state --without normalization--
\begin{equation}
    \ket{000}^{A_a,A_B,A_C} + \ket{020}^{A_a,A_B,A_C} + \ket{112}^{A_a,A_B,A_C} + \ket{121}^{A_a,A_B,A_C}
\end{equation}
and send $A_a$ to $a$, $A_B$ to $B$ and $A_C$ to $C$ using quantum teleportation. Then $B$ create an ancilla $\ket{0}^{B_b}$ and applies the unitary which achieve the transformation
\begin{subequations}
    \begin{align*}
        \ket{00}^{B,B_b}\rightarrow \ket{00}^{B,B_b}\\
        \ket{10}^{B,B_b}\rightarrow \ket{10}^{B,B_b}\\
        \ket{20}^{B,B_b}\rightarrow \ket{21}^{B,B_b}\\
    \end{align*}
\end{subequations}
and send $B_b$ to $b$ and $B$ to $D_B$. $C$ create an ancilla $\ket{0}^{C_c}$ and applies the unitary which achieve the transformation
\begin{subequations}
    \begin{align*}
        \ket{00}^{C,C_c}\rightarrow \ket{01}^{C,C_c}\\
        \ket{10}^{C,C_c}\rightarrow \ket{11}^{C,C_c}\\
        \ket{20}^{C,C_c}\rightarrow \ket{20}^{C,C_c}\\
    \end{align*}
\end{subequations}
and send $C_c$ to $c$ and $C$ to $D_C$. The --unnormalized-- state is the following
\begin{equation}
    \ket{000}^{abc}\ket{02}^{D_BD_C} + \ket{011}^{abc}\ket{20}^{D_BD_C} + \ket{100}^{abc}\ket{12}^{D_BD_C} + \ket{111}^{abc}\ket{21}^{D_BD_C}
\end{equation}
Finally, $D$ measures both $D_B$ and $D_C$ with the following --unnormalized-- family of operators
\begin{equation}
    \mathcal{M}_x = \ket{0}^D\bra{02}^{D_B,D_C} + \ket{1}^D\bra{12}^{D_B,D_C} + (-1)^x\ket{0}^D\bra{20}^{D_B,D_C} + (-1)^x\ket{1}^D\bra{21}^{D_B,D_C}
\end{equation}
eventually correct the phase by applying a $Z$ operator on $b$ and send $D$ to $d$, achieving the wanted distribution.
\bibliographystyle{apalike-refs}
\bibliography{References}

\end{document}